\documentclass[%
reprint,
amsmath,amssymb,
pre,aps,
]{revtex4-1}

\usepackage{graphicx}
\usepackage{dcolumn}
\usepackage{bm}
\usepackage{subcaption}
\usepackage{float}

\DeclareMathOperator\erf{erf}

\begin{document}


\title{Estimation of Parameters from Time Traces originating from an Ornstein-Uhlenbeck Process}

\author{Helmut H. Strey}
 \affiliation{Biomedical Engineering Department and Laufer Center for Physical and Quantitative Biology, Stony Brook University, Stony Brook NY 11794-5281.}

\date{\today}

\begin{abstract}
In this article, we develop a Bayesian approach to estimate parameters from time traces that originate from an overdamped Brownian particle in a harmonic potential, or Ornstein-Uhlenbeck process (OU).  We show that least-square fitting the autocorrelation function, which is often the standard way of analyzing such data, is significantly underestimating the confidence intervals of the fitted parameters.  Here, we develop a rigorous maximum likelihood theory that properly captures the underlying statistics.  From the analytic solution, we found that there exists an optimal measurement spacing ($\Delta t = 0.7968 \tau$) that maximizes the statistical accuracy of the estimate for the decay-time $\tau$ of the process for a fixed number of samples $N$, which plays a similar role than the Nyquist-Shannon theorem for the OU-process. To support our claims, we simulated time series with subsequent application of least-square and our maximum likelihood method.  Our results suggest that it is quite dangerous to apply least-squares to autocorrelation functions both in terms of systematic deviations from the true parameter values and an order-of-magnitude underestimation of confidence intervals.  To see whether our findings apply to other methods where autocorrelation functions are typically fitted by least-squares, we explored the analysis of membrane fluctuations and fluorescence correlation spectroscopy.  In both cases, least-square fits exhibit systematic deviations from the true parameter values and significantly underestimate their confidence intervals.  This fact emphasizes the need for the development of proper maximum likelihood approaches for such methods. In summary, our results have strong implications for parameter estimation for processes that result in a single exponential decay in the autocorrelation function.  Our analysis can directly be applied to single-component dynamic light scattering experiments or optical trap calibration experiments. 

\end{abstract}

\maketitle

\section{Introduction}
In many areas of science the experimental data is measured in the form of traces, often as a function of equally spaced time and/or space.  A very common way of analyzing such data is to calculate the autocorrelation function of the data with respect to time and/or space and then to least-square fit the autocorrelation to the expected theoretical model.  The reason for this analysis procedure is convenience: (1) theory predominantely predict dynamic system behavior (especially for stochastic systems) in terms of correlation functions; (2) correlation functions are easily obtained by hardware correlators that dynamically calculate them from streams of photon counts; and (3) correlation functions are much more ecomomical in terms of data storage than raw traces in time and/or space.  The last resaon is mostly historical, since hard-disk and memory storage was expensive in the advent of dynamic light scattering.  Examples for this kind of experiment are: Dynamic light scattering \cite{RN45}, Fluorescence correlation spectroscopy \cite{RN29}, measurement of membrane fluctuations \cite{RN21,RN10,RN36,RN35}, and the calibration of optical traps \cite{RN39}.
Here we would like to address the question on how to properly extract information from such trajectories.  We will show that the analysis procedure of least-square fitting is problematic in the sense that the parameter's confidence intervals of least-square fitting are often orders of magnitude too small.  In particular, we will develop a proper maximum likelihood framework for the Ornstein-Uhlenbeck process \cite{RN28} to estimate the probability distributions of the parameters given the experimental data.  The Ornstein-Uhlenbeck process is equivalent to a problem from statistical physics that is often used in single molecule/particle experiments: an overdamped Brownian particle in a harmonic potential with Hook's constant k and friction coefficient $\gamma$.  The Langevin equation for such a system can simply be written by:
\begin{equation}
\dot x =  - \frac{k}{\gamma }x + \frac{1}{\gamma }f(t)
\label{model}
\end{equation}
where k is the spring constant, gamma is the friction coefficient and f(t) is a randomly fluctuating force.  The solution to this Langevin equation is given by:
\begin{equation}
\left\langle {x(0)x(t)} \right\rangle  = \frac{{k_B}T}{k} \exp \left( { - \frac{k}{\gamma }t} \right)
\label{corrfct}
\end{equation}
The amplitude of the fluctuation is determined by the equipartition theorem, and the relaxation time is determined by the ratio of friction and spring constant.  For convenience, let us rewrite eq.\ref{corrfct} using the mean-square-amplitude $A=\langle {x^2} \rangle=k_{B}T/k$ and the relation time $\tau=\gamma/k$:
\begin{equation}
\left\langle {x(0)x(t)} \right\rangle  = A \exp \left( {- \frac{t}{\tau}} \right)
\end{equation}
Alternatively, when thinking of a particle that is traped in a harmonic potential, we can express the autocorrelation in terms of the amplitude $A$ and the diffusion coefficient of the particle $D=k_{B}T/\gamma$ as:
\begin{equation}
\left\langle {x(0)x(t)} \right\rangle  = A \exp \left( {- \frac{D}{A}}t \right)
\end{equation}
We will see later that the choice of parameters $(A,\tau)$ vs $(A,D)$ makes a difference in terms of parameter estimation from data, even though given the parameters, equivalent pairs of parameters will result in the same data.\\
Us and others have analyzed such time trajectories by calculating the autocorrelation function and then fitting it to an exponential function to extract the spring constant k and the friction coefficient gamma by a least square fit.  A different approach to solving this particular physics problem is by using a probabilistic approach.  This is done by solving the Smolukowski equation for an overdamped Brownian particle in a harmonic potential, as first reported by Ornstein and Uhlenbeck in 1930 \cite{RN28}
\begin{eqnarray}\label{OUp}
p\left( {x,\Delta t\left| {x_0} \right.} \right) =&& \frac{1}{{\sqrt {2\pi A(1-B^{2}(\Delta t))} }}\nonumber\\
&&\times
\exp \left( { - \frac{{{{\left( {x - {x_0}B(\Delta t)} \right)}^2}}}{{2A(1-B^{2}(\Delta t))}}} \right)
\end{eqnarray}
with $B(\Delta t) = \exp \left( { - \frac{\Delta t}{\tau}} \right) = \exp \left( { - \frac{D\Delta t}{A}} \right)$.
As expected, at long $\Delta t$ this distribution is a Gaussian with variance $A$.  This expression is consistent with an autoregressive (AR) model in the field of signal processing \cite{RN50} for small $\Delta t$.
\section{Autocorrelation Function Analysis}
\begin{figure}
  \includegraphics[width=1\linewidth]{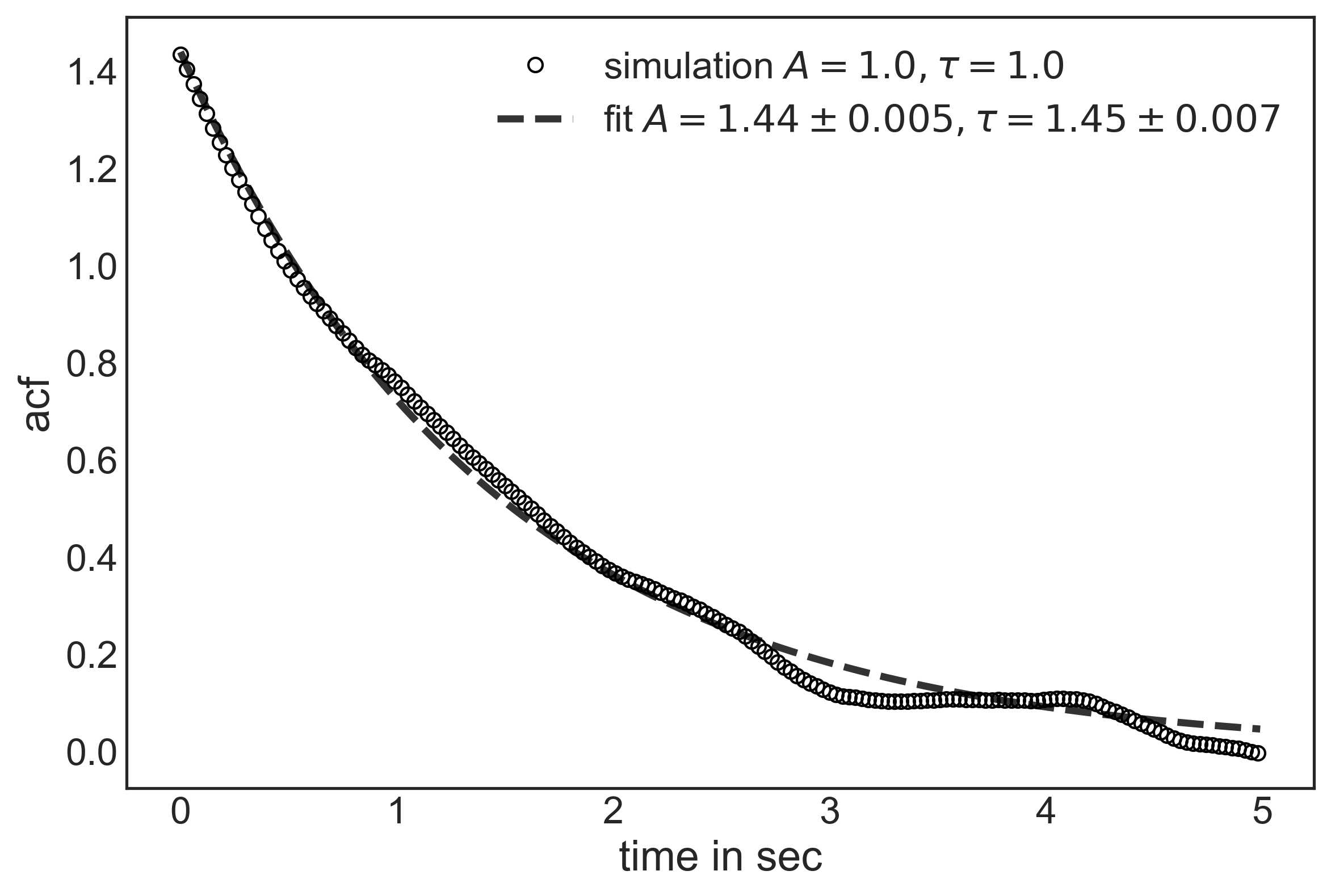}
  \caption{Autocorrelation function of a simulated particle in a harmonic potential with parameters: $A=\langle {x^2} \rangle = 1$ and $\tau=1$. The non-linear least square fit to the data used the Levenberg-Marquardt algorithm \cite{RN41}.}
  \label{fig:acf10000}
\end{figure}
To illustrate the problem with least-square fitting auto-correlation functions (acf), we simulated 10,000 points of an overdamped particle in a harmonic oscillator using the conditional probabilities from eq. \ref{OUp}.  The parameters for the simulation were $A=\langle {x^2} \rangle = 1$, $\tau=1$, and $\Delta t=0.01 sec$.
We calculated the correlation function by Fast Fourier Transformation \cite{RN41}.  Fig. \ref{fig:acf10000} shows a decaying acf and the non-linear least-square fit to an exponential function that was picked from a set of 1000 such simulations that will be presented later in this paper.  This particular fit resulted in an amplitude of $A=1.44\pm0.005$ and a decay time of $\tau=1.45\pm0.007$.  Both parameters are reasonable given that we simulated a time-series that covered just 100 decay times, but the standard deviations of the parameters that resulted from the fits are extremely optimistic: the fitted amplitude deviates from the correct one by 88 standard deviations and the fitted decay time is off by a factor of 64!  To put this in perspective, the probability that the measured amplitude is more than 88 standard deviations away from the real one is $p=1\cdot 10^{-1684}$ - none of our simualted datasets should have ever resulted in such a large deviation.  The question here is, why does a non-linear least square fit underestimate the confidence intervals by so much?
The first evidence comes from a visual inspection of the fit.  In Fig. \ref{fig:acf10000}, the residuals are highly correlated in time which is not surprising given that the acf was calculated from a continuous time series that is correlated in time \cite{RN17}.  What is concering to us is that in many publications the parameters of a least-square fit to a correlation function are taken at face value including their estimated confidence intervals.  This is especially troubling when considering the reproducibility crisis in the life sciences \cite{RN44}.  Because the confidence interval that results from a least-square fit is so small, researchers may not repeat experiments that are expensive and therefore publish results that may not be reproducible.\par

Recognizing the problems with least-square fits several attempts have been made to correct this problem by developing appropriate weights for the least-square fit \cite{RN11,RN32}.  In a least-square fit each data point is assumed to be independent and to be drawn from a Gaussian distribution with a standard deviation $s_{i}$.  In our analysis we assumed that all $s_{i}$ are equal, but in reality, different parts of the acf exhibit different variances.  Unfortunately, this is a circular argument since the parameters have to be known in advance to estimate the errors as function of $\Delta t$.  In the next few sections, we will show a more detailed analysis of the statistics of such least-square fits and compare them to the correct Maximum Likelihood estimate.

\section{Bayesian Analysis}
Let us first review the likelihood function and Bayes theorem (see for a comprehensive dicussion \cite{RN42}).  The likelihood function describes the probability that a particular trajectory $\left\{ {{x_1}({t_1}),{x_2}({t_2}), \ldots ,{x_N}({t_N})} \right\}$ originated from a particular model (eq. \ref{model}) with a particular choice of parameters $A$ and $B$ (or $D$).  
The corresponding likelihood function for a specific time trace $\left\{x_i(i\Delta t)\right\}$ is:
\begin{eqnarray}
	&&p\left( \left\{x_i(t_i)\right\} \left| B, A \right.\right) =
	\frac{1}{\sqrt {2 \pi A} }
	\exp \left( { - \frac{{x_1}^2}{2A}}\right)\nonumber\\
	&&\times\frac{1}{{\sqrt {2\pi A(1-B^{2}(\Delta t))}^{(N-1)} }}\nonumber\\
	&&\times\exp \left( { - \sum\limits_{i=1}^{N-1}\frac{{{{\left( {x_{i+1} - {x_i}B(\Delta t)} \right)}^2}}}{{2A(1-B^{2}(\Delta t))}}} \right)
\end{eqnarray}
This is assuming that we do not know anything about the initial state of the oscillator.  Now Bayes theorem tells us that we can reverse the order of the conditional probability to obtain the posterior probability distribution of our model parameters given our data:
\begin{equation}
	p\left( B, A  \left| \left\{x_i(t_i)\right\} \right. \right) = 
	\frac {p\left( \left\{x_i(t_i)\right\} \left| B, A \right.\right)p\left( B,A \right)}
	{p\left( \left\{x_i(t_i)\right\}\right)}
\end{equation}
The second term of the numerator is called the prior and contains all the information about the model parameters before we look at the data.  For example, the fact that $\tau$ and $A$ are positive is information that we can express using the prior.  Without prior knowledge, one typically uses weakly informative or constant priors so as not to bias the data.  The prior is also a way for us to include information from previous data analysis.  For example, if we have many short traces of data, we could use the posterior of the previous analysis as the prior for the next trace.  It is important to understand that if we want to estimate parameters from a measurement we need to look at the posterior probabilities since it is the probability distribution of the parameters (or model) rather than the data (as in the likelihood).  If one assumes constant prior distributions, then the likelihood turns into the posterior by normalizing it with respect to the parameters which is done by the denominator in Bayes theorem.  Often, it is enough to find the maximum of the posterior and the width of this maximum to estimate the parameters of the model given the data. This procedure does not require normalization and is typically called "Maximum Likelihood".  The name is somewhat misleading since we are taking the maximum of the posterior which happens to be proportional to the likelihood for a constant prior.
In the next section we will follow the "Maximum Likelihood" procedure to determine the most likely model parameters and their corresponding uncertainties.
\section{Maximum Likelihood using $A$ and $\tau$ as parameters}\label{mlAt}
In this section we will express the likelihood in terms of $A$ and $B$ assuming uniform priors for both.  Later we will explore more appropriate choices for priors.
\begin{eqnarray}
	&&p\left( \left\{x_i(t_i)\right\} \left| B, A \right.\right) \propto
	\frac{1}{\sqrt {2 \pi A}^{N} }
	\frac{1}{{\sqrt {(1-B^{2})}^{(N-1)} }}\\
	&&\times\exp \left( -\frac{1}{2A}\left({ {x_1}^{2} + \sum\limits_{i=1}^{N-1}\frac{{{{ {x_{i+1}^{2} - 2x_{i+1}{x_i}B} +{x_i}^{2}B^{2} }}}}{{(1-B^{2})}}} \right)\right)\nonumber
\end{eqnarray}
In order to find the maximum likelihood it is convenient to take the logarithm of $p$ and then take the derivatives with respect to $\sigma$ and $B$.
\begin{equation}
\begin{aligned}
	\Phi &= ln \left( p\left( \left\{x_i(t_i)\right\} \left| B, A \right.\right) \right)\\\\
	&= C - \frac{N}{2} ln(A) - \frac{N-1}{2}ln \left( 1-B^{2}\right) -\frac{1}{2A}Q(B)
\end{aligned}
\end{equation}
with
\begin{equation}
	\begin{aligned}
	&Q(B) = {x_1}^{2} + \sum\limits_{i=1}^{N-1}\frac{ {x_{i+1}^{2} - 2x_{i+1}{x_i}B} +{x_i}^{2}B^{2} }{(1-B^{2})}\\
	&= \frac{x_{1}^{2}+x_{N}^{2}}{1-B^2}+\frac{1+B^2}{1-B^2}\sum\limits_{i=2}^{N-1}x_{i}^{2}-\frac{2B}{1-B^2}\sum\limits_{i=1}^{N-1}x_{i}x_{i+1}
	\end{aligned}
\end{equation}
with $C$ representing an unimportant constant. $Q(B)$ reveals the fundamental statistic - the only terms that exclusively contain the data $\{x_{i}\}$:
\begin{equation}
	\begin{aligned}
		a_{EndPoints}&=a_{EP}=x_{1}^{2}+x_{N}^{2}\\
		a_{SumSquared}&=a_{SS}=\sum\limits_{i=2}^{N-1}x_{i}^{2}\\
		a_{Correlation}&=a_{C}=\sum\limits_{i=1}^{N-1}x_{i}x_{i+1}\\
	\end{aligned}
\end{equation}
The derivative with respect to $A$ determines $A_{max}$:
\begin{equation}\label{partialsigma}
	\begin{aligned}
	\left.\frac{\partial}{\partial A}\Phi\right|_{A_{max},B_{max}} &= -\frac{N}{2A_{max}} +\frac{1}{2A_{max}^{2}}Q(B_{max})=0\\
	A_{max} &= \frac{Q(B_{max})}{N}
	\end{aligned}
\end{equation}
Similarly, we can derive an equation to determine $B_{max}$:
\begin{equation}
	\begin{aligned}
	\frac{\partial}{\partial B}\Phi &= \frac{(N-1)B}{1-B^{2}} -\frac{1}{2A}\frac{\partial}{\partial B}Q(B)\\
	\frac{\partial}{\partial B}Q(B) &=  \frac{2}{(1-B^{2})^{2}}\left(Ba_{EP}+2Ba_{SS}-(1+B^{2})a_{C}\right)
	\end{aligned}
\end{equation}
thus by using eq \ref{partialsigma},
\begin{eqnarray}\label{partialB}
&&	\frac{Q(B_{max})}{N}\frac{(N-1)B_{max}}{1-B_{max}^{2}} -\left.\frac{1}{2}\frac{\partial}{\partial B}Q(B)\right|_{B_{max}}=0\\
	&&\left(a_{EP}+(1+B_{max}^{2})a_{SS}-2B_{max}a_{C}\right)(N-1)B_{max}\nonumber\\
	&&=B_{max}Na_{EP}+2NB_{max}a_{SS}-N(1+B_{max}^{2})a_{C}
\end{eqnarray}
after collecting the terms we can solve for $B_{max}$
\begin{eqnarray}\label{Bmax2}
&&B_{max}^{3}(N-1)a_{SS}+B_{max}^{2}(2-N)a_{C}\nonumber\\
&&-B_{max}(a_{EP}+(N+1)a_{SS})+Na_{C}=0
\end{eqnarray}
The cubic equation in $B_{max}$ has one real root in the $[0,1]$ interval and two roots outside. $A$ can be calculated by inserting the solution for $B_{max}$ into eq \ref{partialsigma}.

\begin{figure}
  \includegraphics[width=1\linewidth]{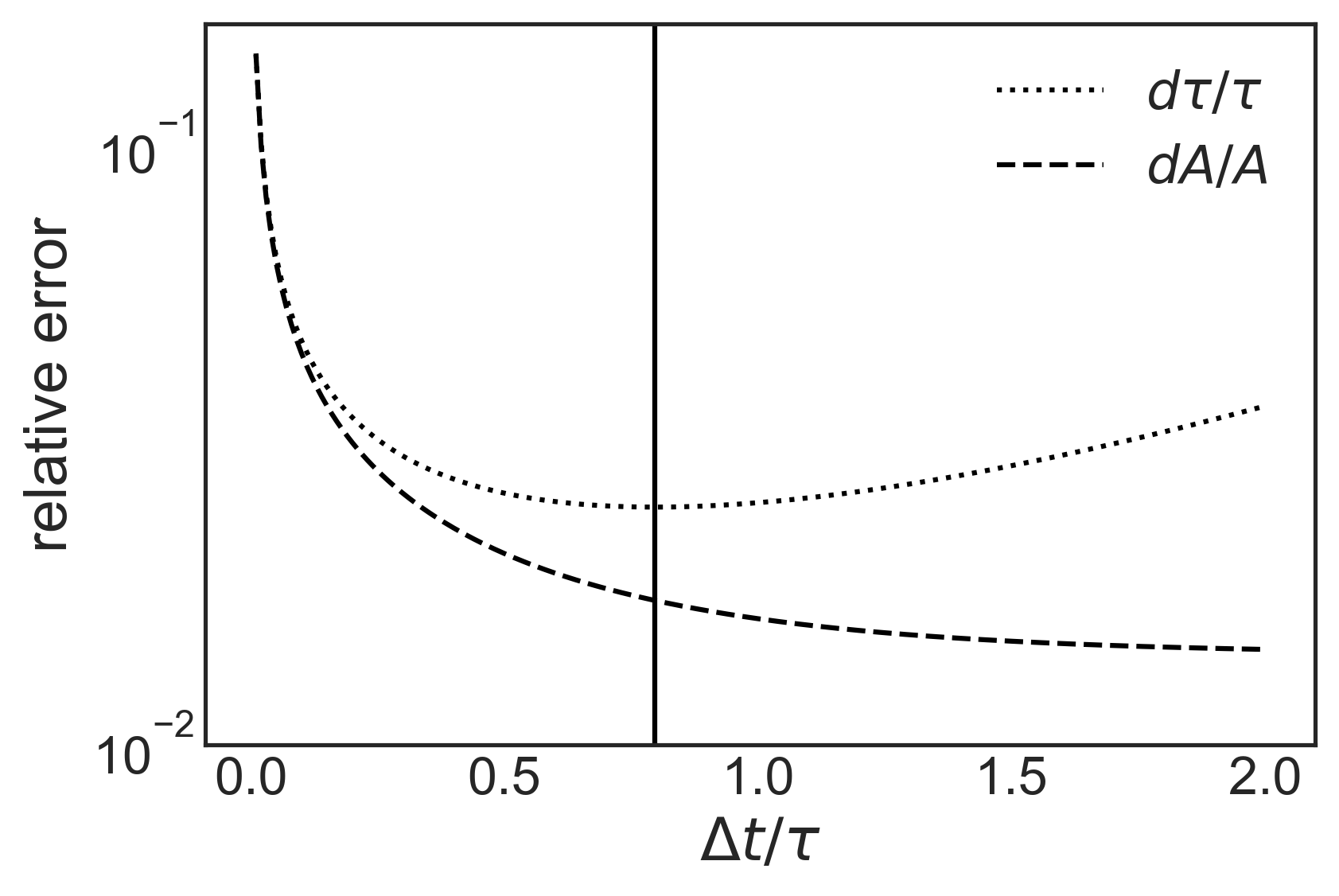}
  \caption{Relative error of $\tau$ and $A$ as a function of $\Delta t/\tau$ at fixed $N=10,000$.  The optimal sampling rate for $\tau$ is at $\Delta t=0.7968\tau$, whereas the relative error of $A$ levels off for larger $\Delta t$.  The vertical red bar indicates the minimum of the relative error of $\tau$, which is independent of $N$ as long as $N$ is large enough}
  \label{fig:optimaldt}
\end{figure}

Next, we need to calculate the uncertainties of the maximum likelihood estimates.  For this we need to calculate the second derivative of the log-likelihood.
\begin{equation}\label{partialsigmasecond}
	\varphi \equiv \left.\frac{\partial^2}{\partial A^{2}}\Phi\right|_{A_{max},B_{max}} = -\frac{N}{2A_{max}^{2}}
\end{equation}
\begin{eqnarray}
	&&\vartheta \equiv \left.{\frac{\partial^{2}}{\partial B^2}}\Phi\right|_{A_{max},B_{max}} =  \frac{(N-1)(1+B_{max}^{2})}{(1-B_{max}^{2})^2}\\
	&&-\frac{(2+6B_{max}^{2})(a_{PP}+2a_{SS})-(12B_{max}+4B_{max}^{3})a_{C}}{2A_{max}(1-B_{max}^{2})^{3}}\nonumber
\end{eqnarray}
\begin{equation}
	\Omega \equiv \left.{\frac{\partial^{2}}{\partial A\partial B}}\Phi\right|_{A_{max},B_{max}} =  \frac{(N-1)B_{max}}{A_{max}(1-B_{max}^{2})}
\end{equation}
In particular, the standard deviation of our model parameters are as follows:
\begin{equation}
	\label{errors}
	\begin{aligned}
		dA_{max}&=\sqrt{\frac{-\vartheta}{\varphi \vartheta - \Omega^{2}}}\\
		dB_{max} &= \sqrt{\frac{-\varphi}{\varphi \vartheta - \Omega^{2}}}\\
		d\tau_{max} &= \frac{\Delta t}{B_{max}\ln^{2}{B_{max}}}dB_{max}\\
	\end{aligned}
\end{equation}
where $\tau_{max} = -\Delta t / \ln{B_{max}}$.\\
To understand the maximum-likelihood confidence intervals better, we calculated the expectation values of $\varphi$, $\vartheta$, and $\Omega$ in the limit of infinitely long datasets: $\left\langle a_{EP} \right\rangle=2A$,$\left\langle a_{SS} \right\rangle=(N-2)A$, and $\left\langle a_{C} \right\rangle=(N-1)AB$.  Inserting this into the previous expressions for $\phi$, $\theta$, and $\Omega$, and collecting only the highest powers of N, we find:
\begin{equation}
	\begin{aligned}
	\left\langle\varphi \right\rangle &= -\frac{N}{2A_{max}^{2}}\\
	\left\langle\vartheta \right\rangle &\approx -N\frac{1+B_{max}^{2}}{(1-B_{max}^{2})^2}\\
	\left\langle\Omega \right\rangle &\approx \frac{NB_{max}}{A_{max}(1-B_{max}^{2})}\\
	\end{aligned}
\end{equation}
Only collecting the highest orders in $N$, we get
\begin{equation}
	\begin{aligned}
	dA_{max} &= \frac{A_{max}}{\sqrt{2N}}\sqrt{\frac{1+B_{max}^{2}}{1-B_{max}^{2}}}\\
	dB_{max} &= \sqrt{\frac{1-B_{max}^{2}}{N}}
	\end{aligned}
\end{equation}
As expected, both confidence intervals $dA_{max}$ and $dB_{max}$ reduce as $1/\sqrt{N}$.  We can now ask the following question: Given that we want to perform $N$ measurement to get an estimate for $\tau$, where $N$ is large, can we find an optimal $\Delta t$ to maximize the statistical accuracy.  We find this optimal $\Delta t$ by minimizing $d\tau_{max}/\tau$ with respect to $\Delta t/\tau$.
\begin{equation}
	d\tau_{max}/\tau = \frac{\sqrt{\exp(2\Delta t/\tau)-1}}{\Delta t/\tau}
\end{equation}
We determined the minimum numerically at $\Delta t = 0.7968\tau$ as shown in Fig. \ref{fig:optimaldt}.  The optimal sampling frequency for $\tau$ plays a similar role than the Nyquist-Shannon sampling theorem, with the difference that the latter gives a limit for the frequency that can be sampled, whereas the former denotes the "best" frequency to sample to obtain the most accurate result at fixed number of sampling points $N$.  In our case any sampling frequency will measure both the amplitude $A$ and $\tau$ at a sufficiently large enough sample size. In the same figure, we also show how the relative error of the amplitude $A$ changes with $\Delta t/\tau$.  When sampling at high frequency (at $\Delta t < \tau$) the samples are not fully independent, and therefore one needs more samples to achieve the same accuracy.  At larger sampling steps ($\Delta t >> \tau$) the samples become independent and the accuracy is only determined by the number of samples $N$.  Knowing the behavior of sampling frequency on accuracy enables experimental designs that optimze the sampling frequency as the measurement are taken.  For example, one could chose the sampling frequency according to the importants of knowing $\tau$  or $A$ and their standard deviations.  Another design could be to measure a small sample at a fixed frequency, analyze that sample and extract first estimates for $A$,$dA$,$\tau$ and $d\tau$ and then change the sampling frequency to the optimal one.  This is especially important when data is expensive.  For example, in financial markets it is typically more costly to obtain higher frequency trading data, and our analysis could reveal where the optimal cost to benefit lies.
Our analytic results are exact for data sets of arbitrary length, given that the maximum likelihood method is appropriate.  For long data sets our analysis converges to a similar analytical approach that was taken in \cite{RN51} for use in optical trap calibrations. Our method is also significantly easier to implement than the maximum likelihood approach using the Powerspectrum of the data as developed in \cite{RN52}, even though we believe that the two methods give the same results. So far we assumed that the mean of $x$ is zero. Our theory can be extended easily to include a parameter $\bar{x}$ to independently estimate the value around which $x$ fluctuates.
\section{Comparison and Validation using Simulation}
We will now compare our analytic results with simulated time-series to see how the true posterior parameter estimate compares to the exponential least-square fit.  The first thing that becomes immediately obvious from the maximum likelihood expressions is that the parameter estimates only depend on the first two points of the autocorrelation function ($a_{SS}$ and $a_{C}$) given that the time-series is very long and that the first and last point can be neglected (as given by the fundamental statistics).  This means that least-square fitting the whole auto-correlation function, while tempting, is misleading and unnecessary.  In fact, it seems that trying to fit the fluctuations of the tail of the autocorrelation function may very well be the reason why this approach gives the wrong result.\par
We simulated data sets of amplitude $A=1$, $\tau=1$, and a time step of $\Delta t=0.01$ for different dataset lengths $N$.  A choice of $\Delta t=0.01$ means that we cover each relaxation time with 100 data points which is reasonable from an experimental point of view.
As in Section II, we simulated time series with the chosen parameters $(N,\Delta t)$ using the conditional probability for the Ornstein-Uhlenbeck process.  We then estimated the parameters $A$ and $\tau$ using the analytic solution as well as the exponential least-square fit with constant weights (LS) to the autocorrelation function which we calculated using FFT.  In addition, we performed least-square fits with appropriate weights (LSW) (see \cite{RN11,RN32}).  We first used all acfs to estimate the standard deviation of each point of the acf.  This standard deviation was then used as weights for the non-linear least-square fit. 

\begin{figure}
    \centering
    \begin{subfigure}[b]{0.35\textwidth}
        \includegraphics[width=\textwidth]{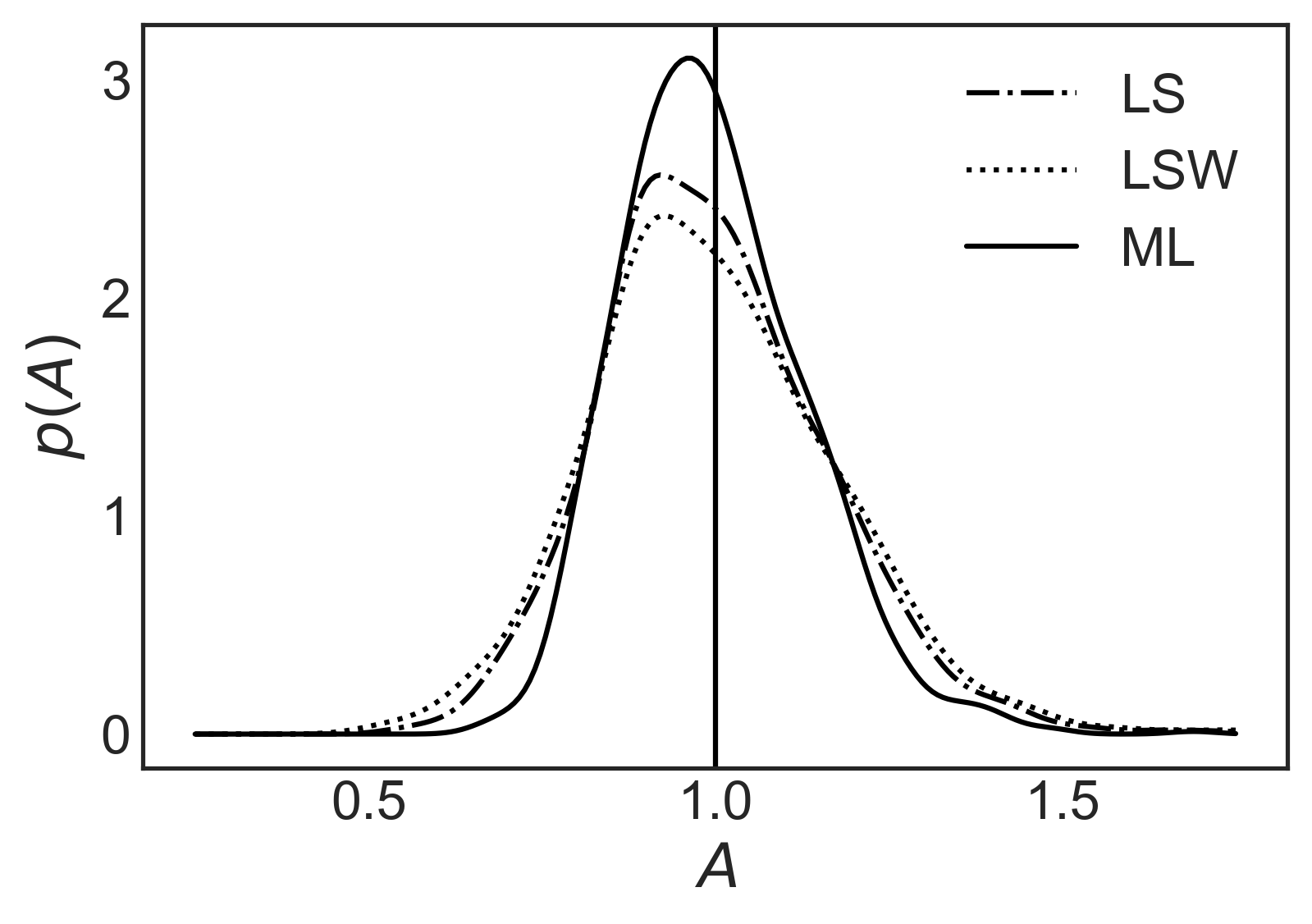}
        \caption{histogram of $A$}
        \label{fig:p_sigma2}
    \end{subfigure}
    ~ 
    \begin{subfigure}[b]{0.35\textwidth}
        \includegraphics[width=\textwidth]{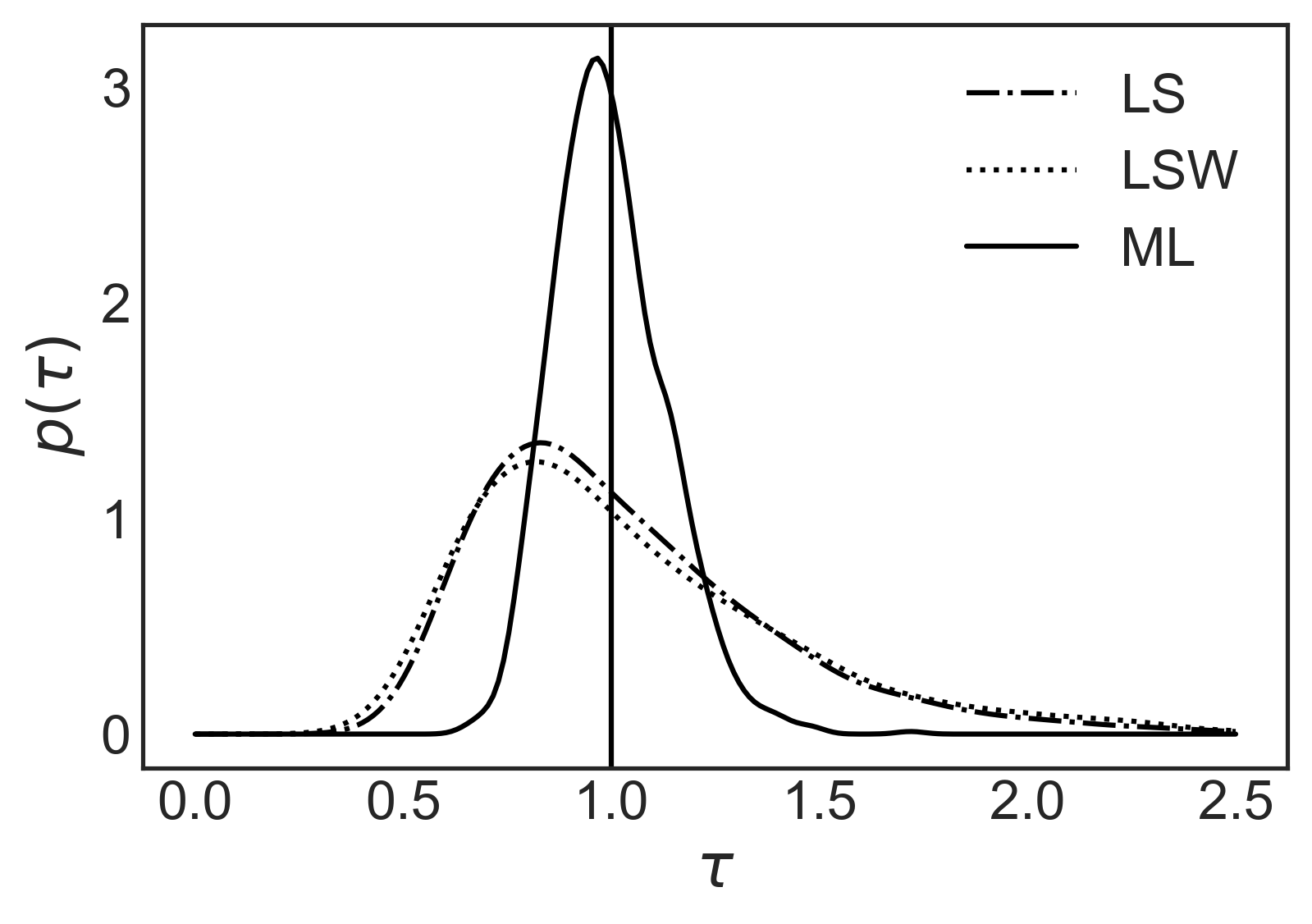}
        \caption{histogram of decay time $\tau$}
        \label{fig:p_tau}
    \end{subfigure}
    \caption{Comparison of probability distributions of $A$ and $\tau$ determined by least square fit and Maximum Likelihood Analysis.  We plotted the estimated probability distributions from the samples using Gaussian Kernel Density Estimation (KDE implementation in SciPy: scipy.stats.gaussian\_ked). The least square estimates for $\tau$ are significantely wider than the Bayesian estimate.}\label{fig:p_fig}
\end{figure}

\begin{figure}
    \centering
    \begin{subfigure}[b]{0.35\textwidth}
        \includegraphics[width=\textwidth]{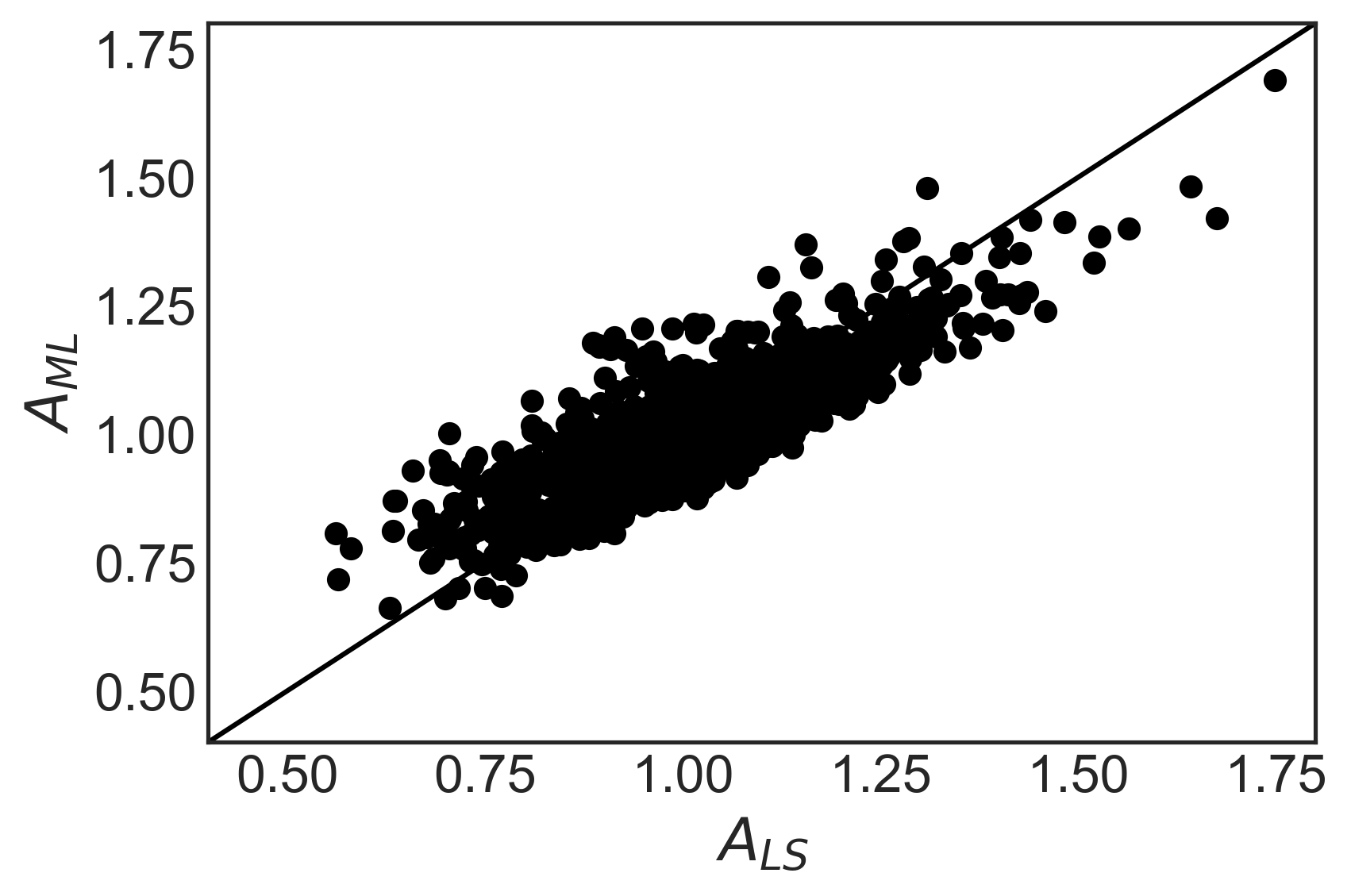}
        \caption{correlation of $A$}
        \label{fig:LS_sigma2}
    \end{subfigure}
    ~ 
    \begin{subfigure}[b]{0.35\textwidth}
        \includegraphics[width=\textwidth]{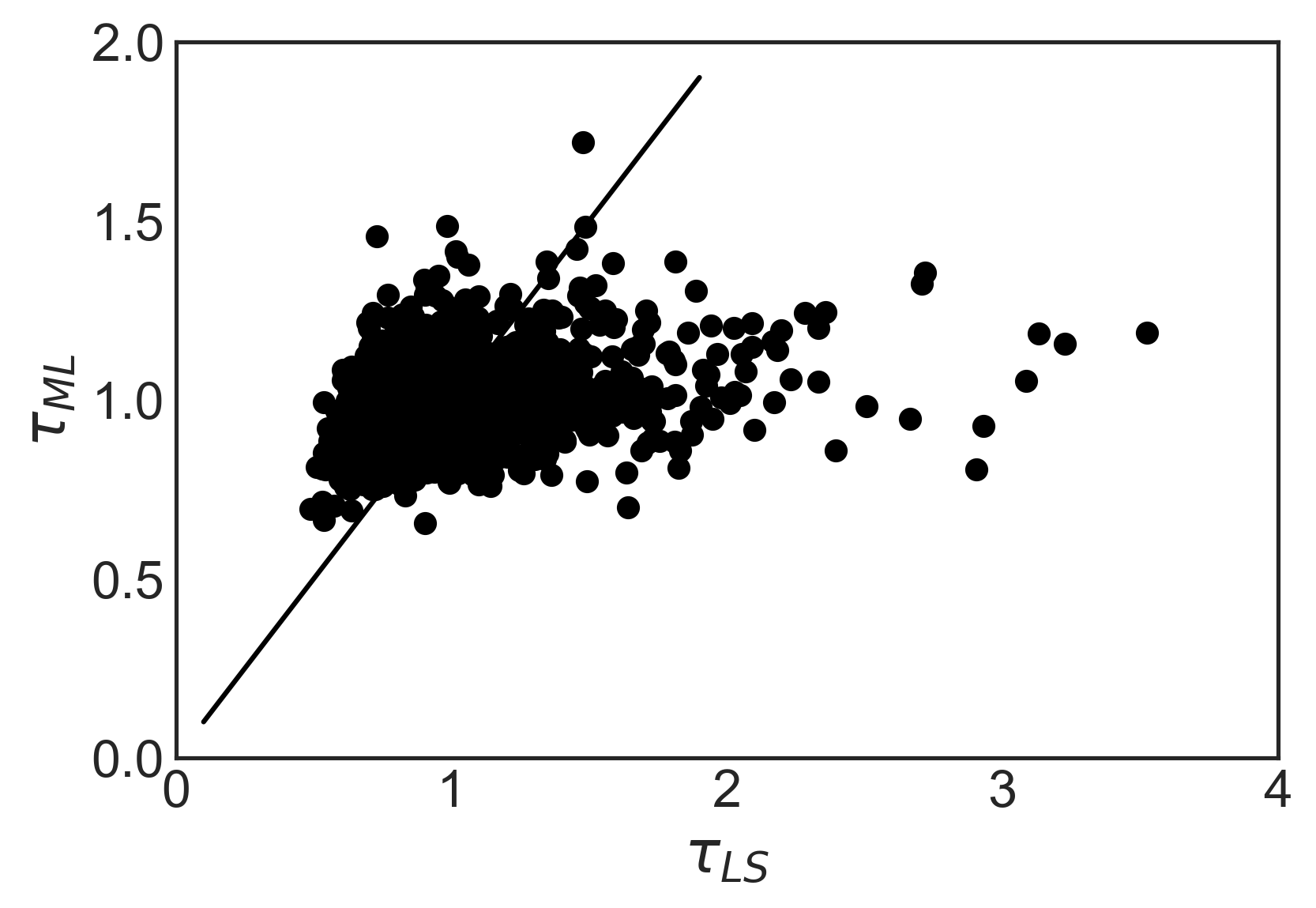}
        \caption{correlation of decay time $\tau$}
        \label{fig:LS_tau}
    \end{subfigure}
    \caption{Correlations between least square estimate of $A$ and $\tau$ to estimates derived from Maximum Likelihood (LS) Analysis.  The $A$ are highly correlated whereas the $\tau's$ seem uncorrelated}\label{fig:LS_fig}
\end{figure}
In Figs. \ref{fig:p_fig} and \ref{fig:LS_fig}, we summerize 1000 simulations of 10,000 points each for $\Delta t=0.01$ which is equivalent of simulating time traces of 100 relaxation times with a resolution of 100 points per relaxation time.  The results suggest that the least square estimate for the amplitude $A$ compares well with the Maximum Likelihood estimate, also indicated by the strong positive correlation between the two methods.  Specifically, the standard deviations of the $A$ distributons are $\sigma_{LS}=0.16$, $\sigma_{LSW}=0.18$, and $\sigma_{ML}=0.13$.  These standard deviations have to be compared to the estimated errors from the fit or maximum likelihood estimate: $dA_{LS}=0.011$, $dA_{LSW}=0.014$, and $dA_{ML}=0.14$. Both least-square fits underestimate the error by a factor of 10, whereas the maxiumum likelihood error estimate is in agreement with the distribution.  On the other hand, least square estimates poorly track the estimates for $\tau$.  Firstly, the probability distribution for the least square estimate for $\tau$ is much wider, and secondly both estimates seem uncorrelated to the Bayesian result.  Specifically, the standard deviations of the $\tau$ distributions are $\sigma_{LS}=0.39$, $\sigma_{LSW}=0.44$, and $\sigma_{ML}=0.13$.  The estimated errors from the fits are $d\tau_{LS}=0.016$, $d\tau_{LSW}=0.018$, and $d\tau_{ML}=0.14$.  Again, the least-square error estimates are significantly underestimating the real distribution, in this case by a factor of 24.  From our results it is also obvious that performing the least-square fits with appropriate weights does not improve the performance.  In some case it actually worsens the estimates.

\begin{figure}
  \includegraphics[width=1\linewidth]{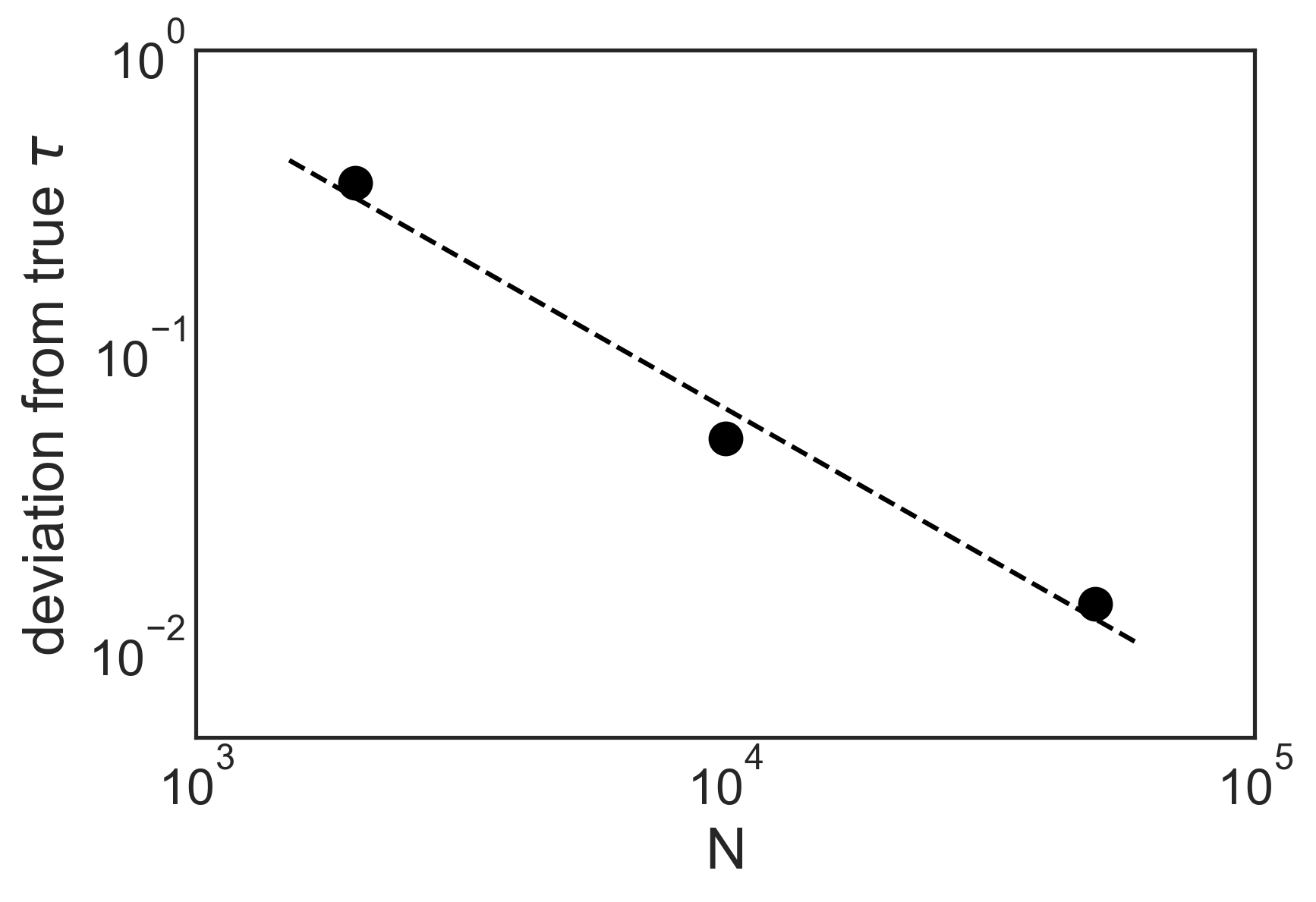}
  \caption{Systematic deviation of the least-square estimate of $\tau_{LS}$ from the true $\tau$ as function of number of data points $N$ for $A=\tau=1$. The fit to the data indicates that the systematic deviation from the true $\tau$ is inversely related to the number of data points.}
  \label{fig:deviation}
\end{figure}

In addition to a significant overestimation of confidence we found that least-square fits also systematically deviate for the estimation of $\tau$.  Fig. \ref{fig:deviation} shows the systematic deviation from the true $\tau$ as function of number of data points $N$ using the same parameters as before ($\Delta t = 0.01$, $A=\tau=1$).  In order to detect this systematic deviation, we simulated 1000 time traces so that after averaging $\tau_{LS}$ the standard error of the mean was significantly less than the deviation from the true mean.  At $N=2000$, we found that at $\tau=1.36\pm0.04$ , at $N=10,000$ we found $\tau=1.05\pm0.004$ and at $N=50,000$ we found $\tau=1.0014\pm0.0001$.  These findings are consistent with that the deviation from the true $\tau$ is inversely related to the number of data points.

\section{Maximum Likelihood using $A$ and $D$ as parameters}
In this section, we explore a different choice of independent parameters.  Instead of writing the likelihood as function of $A$ and $B$ we will express it as
\begin{eqnarray}
	&&p\left( \left\{x_i(t_i)\right\} \left| D, A \right.\right) \propto
	\frac{1}{\sqrt {2 \pi A}^{N} }
	\frac{1}{{\sqrt {(1-B(A,D)^{2})}^{(N-1)} }}\nonumber\\
	&&\times\exp \left( -\frac{1}{2A}Q(B(A,D))\right)
\end{eqnarray}
with $B(A,D)=\exp \left( { - \frac{D\Delta t}{A}} \right)$. In principle, one could follow the maximum likelihood procedure that we employed in section \ref{mlAt}.  Unfortunately, writing the derivatives in analytical form and finding the maximum likelihood values and standard deviations becomes significantly more difficult.  Instead we chose to explore this case using Markov-Chain Monte-Carlo (MCMC) methods.  We formulated the problem in python (Numpy, Scipy, PYMC3\cite{RN75}) to cover the following case: $(A=1,D=1)$.  It is sufficent to investigate just one ratio of $A$ to $D$ since any ratio can be obtained by choosing a different $\Delta t$.  In other words, plotting $dA/A$ and $dD/D$ as a function of $\Delta t/\tau$ will produce the same curve for any ratio $A/D$.

\begin{figure}
    \includegraphics[width=1\linewidth]{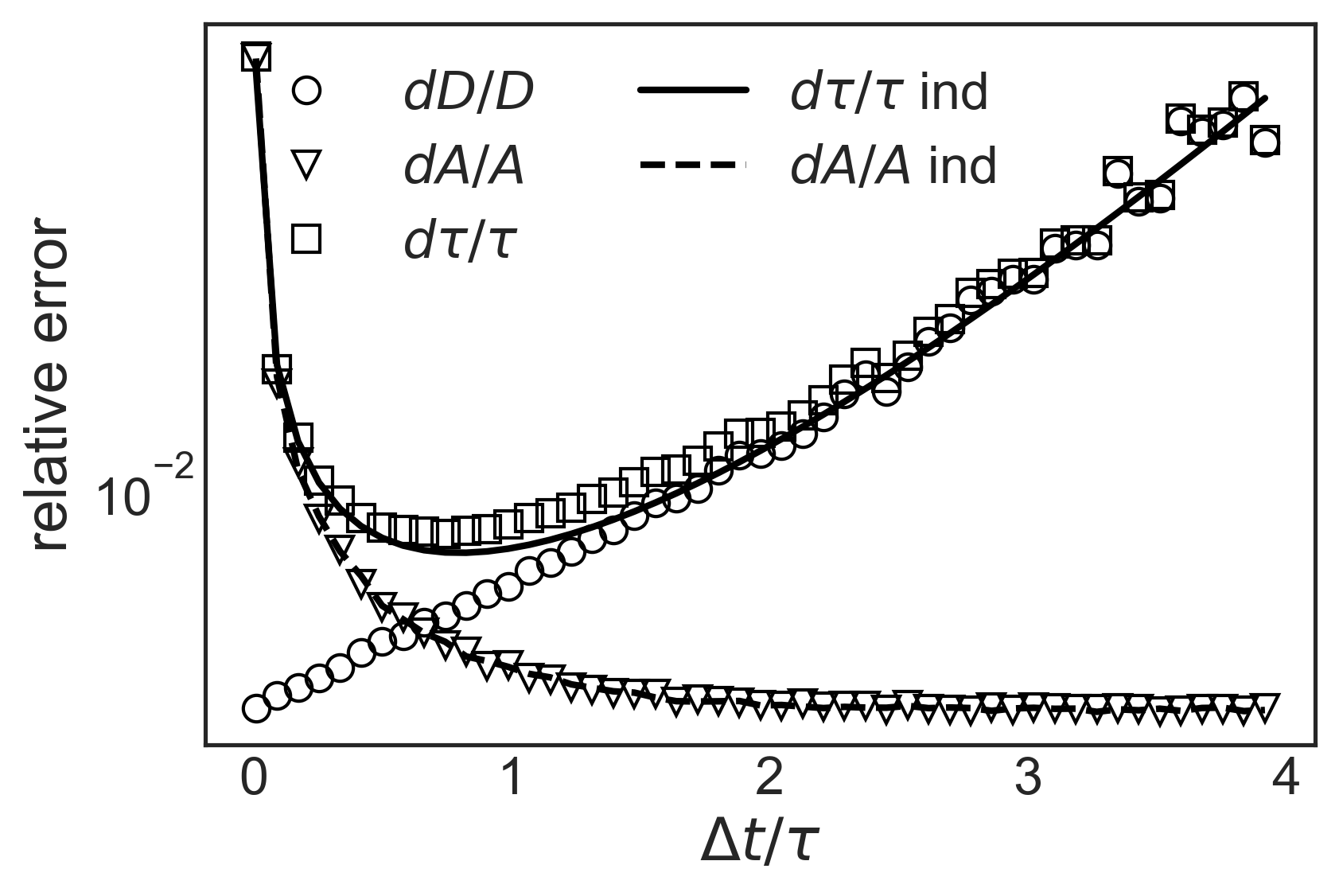}
    \caption{Plots of relative errors of $D$, $A$ as independent variables compared to $A$ and $\tau$ as independent variables as function of $\Delta t$}\label{fig:ad_fig}
\end{figure}

We simulated 50 time traces of 100,000 points with $\Delta t/\tau$ varying from 0.01 to 4.  We analyzed these time traces using PYMC3 assuming constant priors producing 82,000 samples for $A$ and $D$.  From these traces, we calculated the values of maximum likelihood values of $A$ and $D$ and their standard deviations.  Using error progression, we can then calculate the mean value and standard deviation of $\tau$.  Fig.\ref{fig:ad_fig} shows plots of the relative errors of both $A$, $D$ and $\tau$ as function of $\Delta t$.  We also inserted the corresponding analytical results for $A$ and $\tau$ as idependent variables.  For the range of $\Delta t/\tau$ that we considered as resaonable, it seems that picking different sets of independent variable made almost no difference in the behavior of relative errors as function of sampling frequency.
Interestingly, looking at the behavior of $dD/D$ and $dA/A$, the optimal sampling frequency can be explained by opposite trends in $A$ and $D$.  $dA/A$ goes down whereas $dD/D$ goes up linearly as $\Delta t$ increased.  As we explained before, the longer $\Delta t$ is the more independent the data points become which leads to a better estimate for the amplitude.  On the other hand, longer $\Delta t$ reduces the effect of diffusion on the data points.  At the extreme end the conditional probability between data points becomes time independent for $\Delta t\rightarrow \infty$.  In practice, if faced with a situation where $A$ and $D$ are independent variables, our results suggest that the best procedure is to analyze the data as if $A$ and $\tau$ are independent and then to calculate the error by error progression to calculate $dD$.

\section{Practical Applications}
In this section, we will explore more complex scenarios in which least-square fitting autocorrelation function can lead to overestimating the confidence intervals of parameters.  In particular, we will focus on the analysis of measured data from lipid membrane fluctuations and fluorescence correlation spectroscopy.  In both cases the resulting autocorrelation functions are sums or integrals over exponential decays with varying decay times and it is not immediately obvious that our previous results apply to such systems.
\subsection{Analysis of membrane fluctuations}
\begin{figure}
    \centering
    \begin{subfigure}[b]{0.4\textwidth}
        \includegraphics[width=\textwidth]{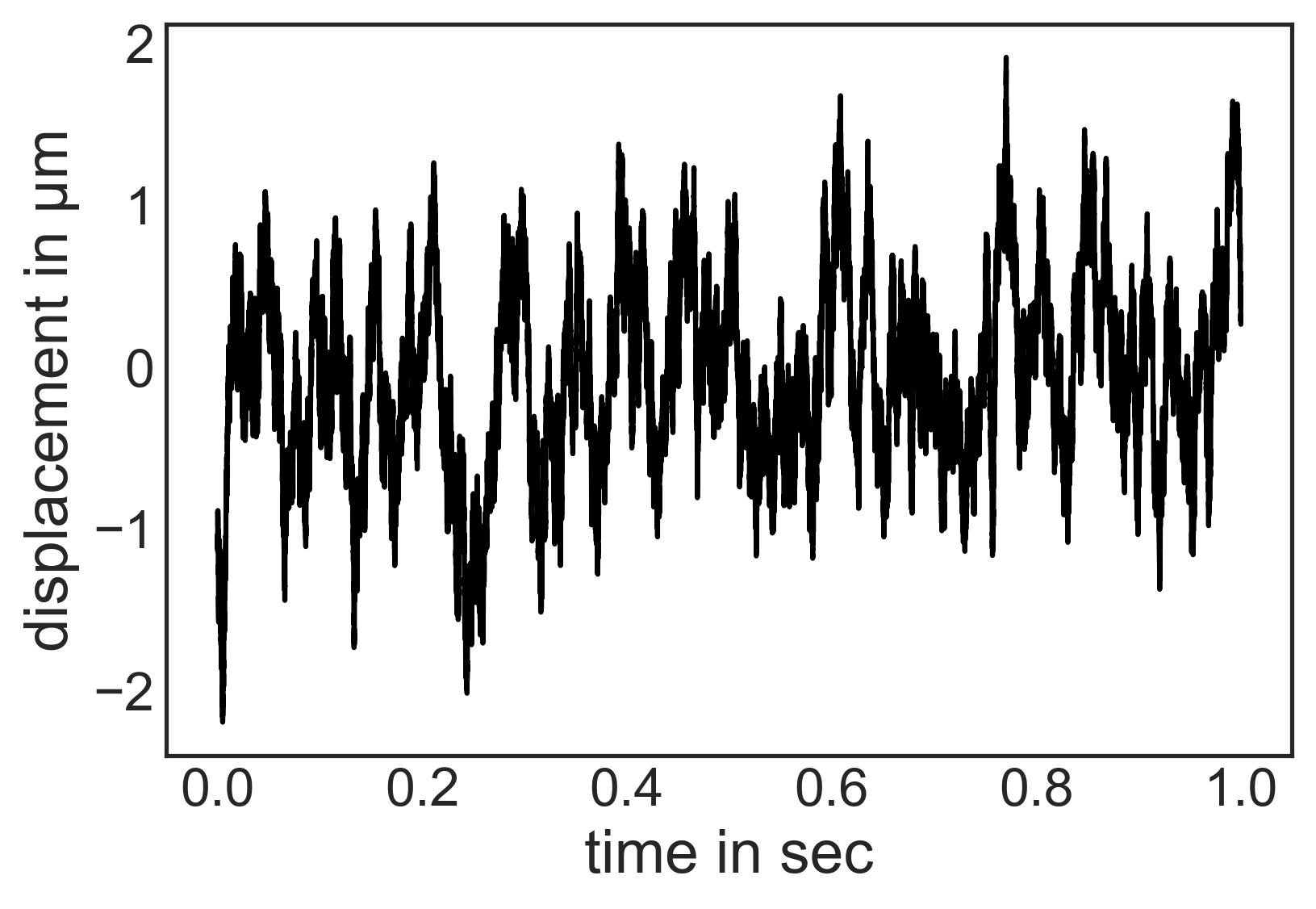}
        \caption{Trace of membrane fluctuations}
        \label{fig:mts}
    \end{subfigure}
    ~ 
    \begin{subfigure}[b]{0.4\textwidth}
        \includegraphics[width=\textwidth]{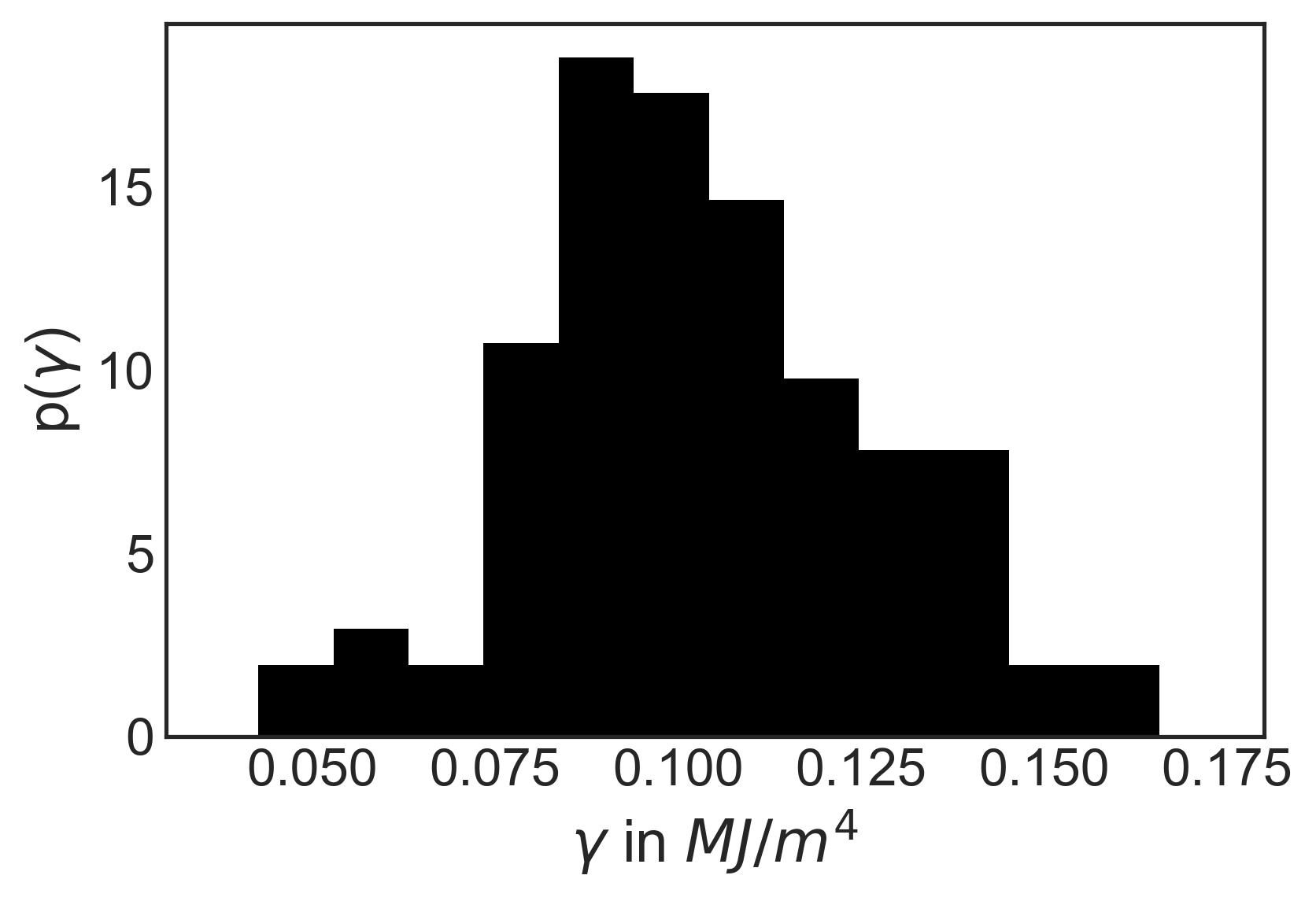}
        \caption{histogram of membrane tension $\gamma$}
        \label{fig:p_gamma}
    \end{subfigure}
    \caption{Results for least-square fitting the autocorrelation functions of simulated membrane flucuations}\label{fig:membrane_fig}
\end{figure}
In \cite{RN21,RN10}, the authors analyze high-resolution lipid membrane fluctuations using a novel technique (dynamic optical displacement spectroscopy) with a $20nm$ height and $10\mu m$ time resolution.  In short, by placing a laser focus across a fluorescent lipid membrane, small bending fluctuations can be measured by changes in fluorescent intensity.  Because this technique measures the fluctuations at one point, the height fluctuations are comprised of an infinite sum of eigenmodes with different wave-vectors.  In particular, the Hamiltonian of an elastic membrane is given by:
\begin{equation}
	H = \int_{S}d^{2}x \left[ \frac{\kappa}{2}(\nabla^{2}h)^{2} + \frac{\sigma}{2}(\nabla h)^{2} + \frac{\gamma}{2}(h-h_{0})^{2} \right]
\end{equation}
from this Hamiltonian we can calculate the autocorrelation function of height fluctuations at a particular point using the fluctuation-dissipation theorem:
\begin{equation}
	\left\langle {\Delta h(x,y,0) \Delta h(x,y,t)} \right\rangle  = k_{B}T \sum_{q^{2}=q_{x}^{2}+q_{y}^2} \frac{\exp(-\Gamma(q)t)}{\kappa q^{4} + \sigma q^{2} + \gamma}
\end{equation}
with
\begin{equation}
	\Gamma(q)=\frac{\kappa q^{4} + \sigma q^{2} + \gamma}{4\eta q}
\end{equation}
comparing this result with eq.\ref{corrfct}, we see that the correlation function of membrane fluctuations is the sum of harmonic oscillators whose spring constant and friction coefficient varies with the wave-vector $q$.  Because we don't have access to original data, we simulated realistic membrane fluctuation data and analyzed it using a least-square fit to the correlation function and campare it to the correct Bayesian model.  Specifically, we assumed the following realistic parameters for the simulation: $T=25^{\circ}C$, $\kappa = 10k_{B}T$, $\sigma=0.5\mu J/m^{2}$, $\gamma=0.1MJ/m^{4}$ and $\eta = 1.0mPa s$.  The minimum wave-vector $q_{min}=\sqrt{3}/10\mu m$ to describe a $10\mu m$ vesicle or red blood cell.  We simulated the modes by assuming that $q_{lm} = q_{min}\sqrt{l^{2}+m^{2}}$ and because of the rapid decline in amplitude and relaxation rate, we only consider $l<4,m<4$ resulting in the simulation of the first 15 modes.  We simulate each mode using the conditional probability for the Ornstein-Uhlenbeck process using a $\Delta t=10\mu s$ for a total time of 1 sec after which we sum all the modes into a single time trace and calcualte the autocorrelation function using FFT.  Similar to \cite{RN21,RN10}, we least-square fitted the autocorrelation function with fixed $\kappa,\sigma, \eta$ to estimate the membrane tension $\gamma$.  Fig. \ref{fig:membrane_fig} (left) shows a simulated time series and (right) shows a histogram of $p(\gamma)$ from 100 simulations which resulted in an $\gamma=0.102\pm0.025MJ/m^{4}$.  This is a reasonable result for the real value of $\gamma=0.1MJ/m^{4}$.
On the other hand the average error estimation that resulted from the least-square fit was $\Delta \gamma =6\cdot 10^{-5}MJ/m^{4}$ which is 400 times smaller than the observed distribution.

To be fair, rather than relying on the error estimates from the least-square fits, the authors of \cite{RN21,RN10} reported confidence of their model parameters derived from repeated measures.  Even though appropriate in this case, we find this procedure unsatisfactory in general since often biological experiment-to-experiment variance is larger than the accuracy of individual experiments.
\section{Fluorescence Correlation Spectroscopy}
\begin{figure}
    \centering
    \begin{subfigure}[b]{0.4\textwidth}
        \includegraphics[width=\textwidth]{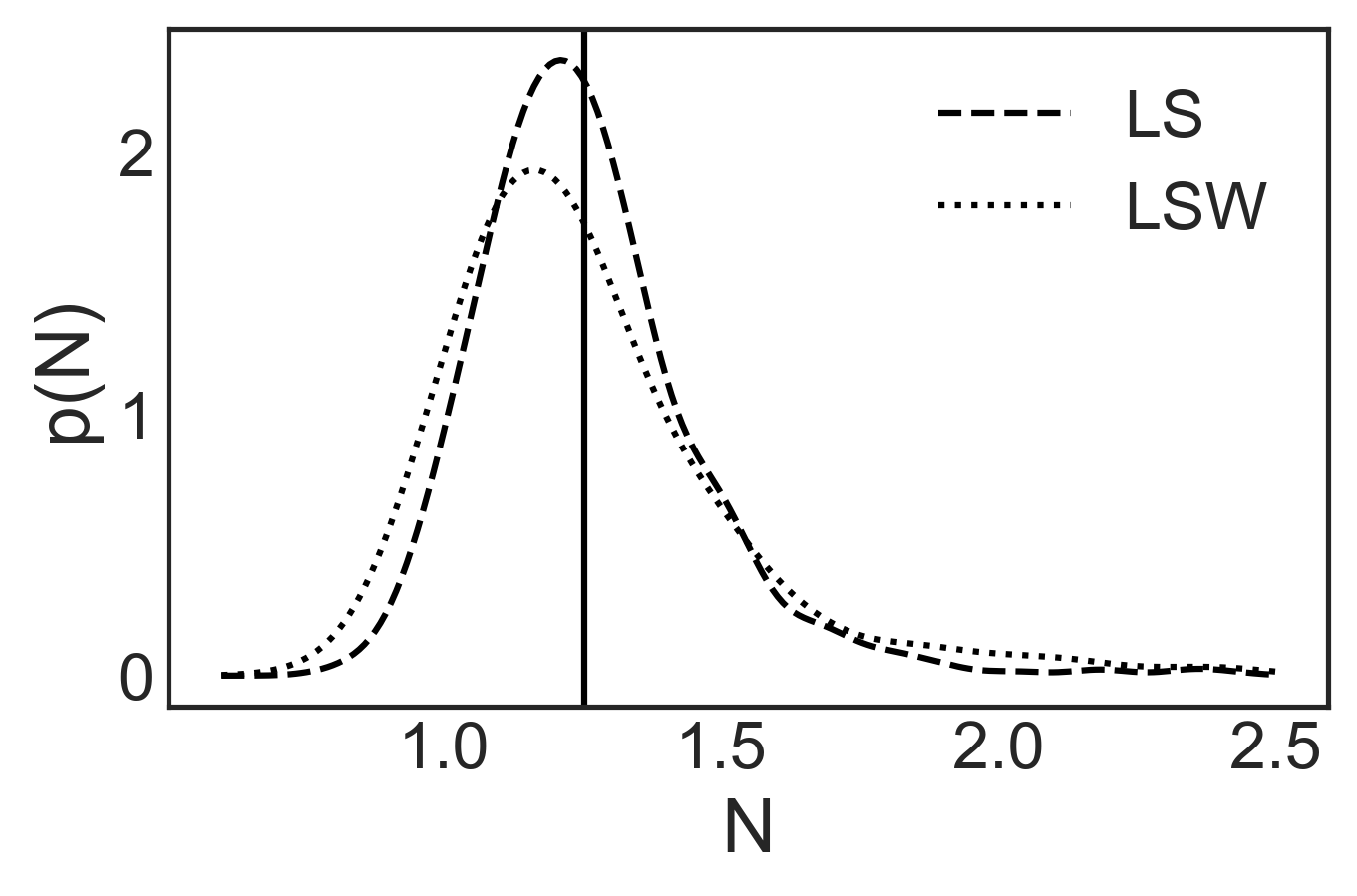}
        \caption{histogram of number of particles $N$}
        \label{fig:p_N_short}
    \end{subfigure}
    \qquad 
    \begin{subfigure}[b]{0.4\textwidth}
        \includegraphics[width=\textwidth]{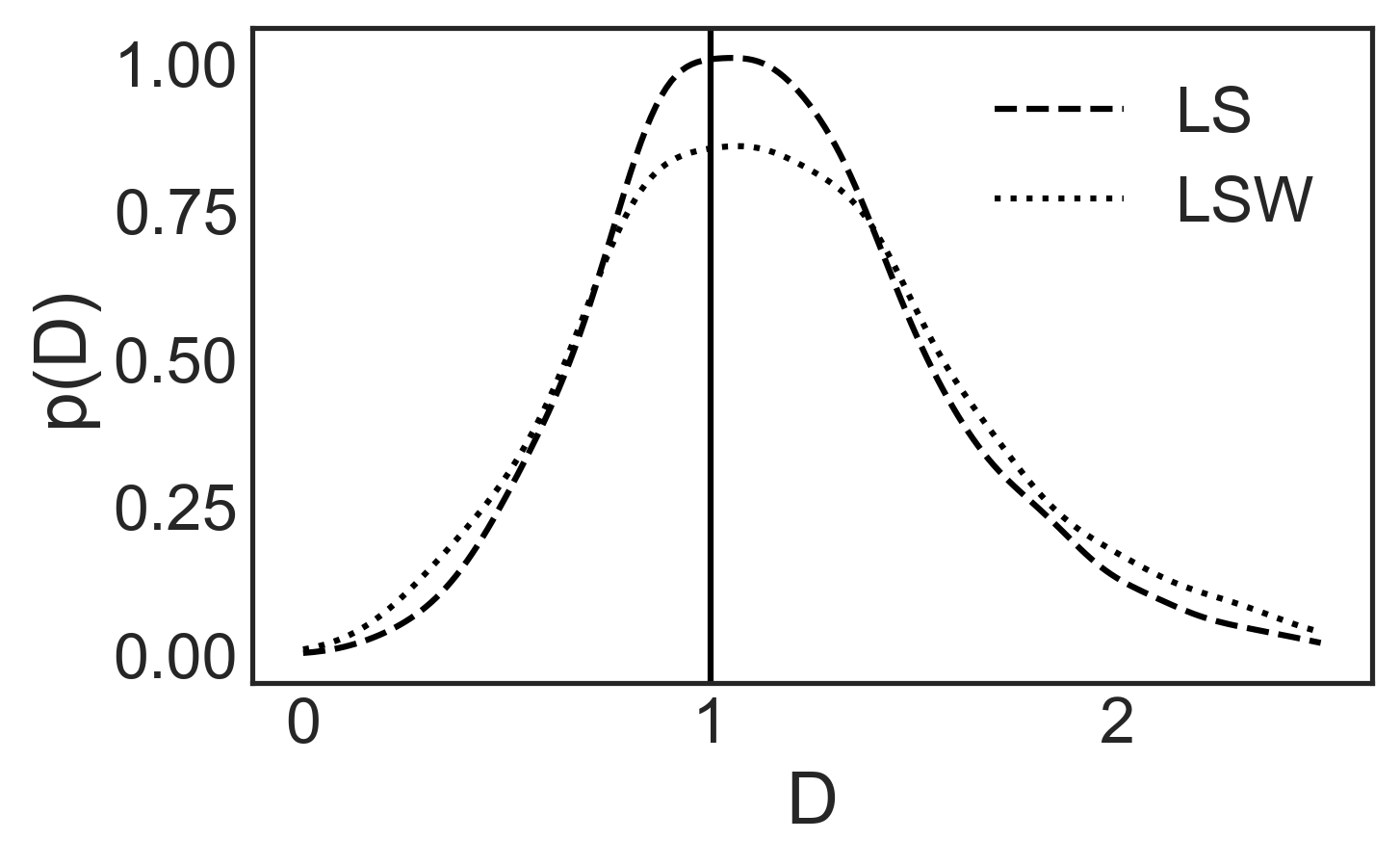}
        \caption{histogram of diffusion constant $D$}
        \label{fig:p_D_short}
    \end{subfigure}
    \caption{Comparison of Gaussian KDE probability distributions of $N$ and $D$ or traces of 10,000 points determined by least square fit with and without weights.}\label{fig:p_ND_short}
	\label{fig:FCSfig_short}
\end{figure}

Fluorescence correlation spectroscopy (FCS) is a powerful technique to measure concentrations and diffusion coefficients of fluorescent molecules \cite{RN29}.  Here we want to illustrate that, similar to a simple Orstein-Uhlenbeck process, fitting of FCS correlation functions leads to a similar underestimation of errors in the fitting parameters.  In order to show this, we are creating artifical and idealized datasets by numerical simulation in one dimension.  In order to simplify the simulation, we are limiting the simulation to a finite box of length $2L$ spanning $[-L,L]$ in which we place N fluorescent particles.  For the simulation we are imposing reflective boundary conditions on the box which means that molecules leaving the box on the right will be reflected back into the box.  Such boundary conditions would be appropriate for fluorescent molecules enclosed in a cell.  In order to calculate the expected autocorrelation function we have to create an intensity function that is periodic with a period of 2L.  We will do this using a Fourier series $a_{n}$ of a Gaussian Intensity distribution.
\begin{equation}
\phi(x)=I_{0}\exp{\left(- \frac{2x^{2}}{w^2}\right)}
\label{illpro}
\end{equation}
The autocorrelation function $I(t)I(0)$ can then be described as:
\begin{eqnarray}
&&\left\langle {I(0)I(t)} \right\rangle=
\int_{-\infty}^{\infty}dx_{1}\int_{-L}^{L}dx_{2}\nonumber\\
&&\times\frac{1}{\sqrt{4\pi Dt}}\exp{\left(- \frac{\left(x_{1}-x_{2}\right)^{2}}{4Dt}\right)}\phi(x_{1})\phi(x_{2})
\end{eqnarray}
where $\phi(x)$ is the illumination profile that is periodic and even with a period of $2L$.  In particular we construct $\phi(x)$ by a Fourier series so that:
\begin{equation}
\phi(x)=\frac{a_{0}}{2} + \sum_{n=1}^{\infty}a_{n}\cos{\frac{\pi nx}{L}}
\end{equation}
with
\begin{eqnarray}
a_{n}=&&\frac{I_{0}}{2L}\int_{-L}^{L}dx\exp{\left(- \frac{2x^{2}}{w^{2}}\right)}\cos{\frac{\pi nx}{L}}\nonumber\\
=&&\frac{I_{0}\sqrt{2\pi w^{2}}}{8L}
\exp{\left(-\frac{n^{2}\pi^{2} w^{2}}{8L^{2}}\right)}\\
&&\times\left(\erf\left(\frac{4L^{2}-in\pi w^{2}}{2\sqrt{2}L w}\right)+\erf\left(\frac{4L^{2}+in\pi w^{2}}{2\sqrt{2}L w}\right)\right)\nonumber
\end{eqnarray}
We can now calculate the autocorrelation function by first integrating over $x_{1}$.
\begin{eqnarray}
&&\int_{-\infty}^{\infty}dx_{1}\frac{1}{\sqrt{4\pi Dt}}\exp{\left(- \frac{\left(x_{1}-x_{2}\right)^{2}}{4Dt}\right)}\cos{\frac{\pi nx_{1}}{L}}\nonumber\\
&&= \exp{\left(-\frac{Dn^{2}\pi^{2}t}{L^{2}}\right)}\cos{\frac{\pi nx_{2}}{L}}
\end{eqnarray}
The autocorrelation function then can be written as:
\begin{equation}
\begin{aligned}
\left\langle {I(0)I(t)} \right\rangle &= I_{0}^{2}\int_{-L}^{L}dx_{2} \left( \frac{a_{0}}{2} + \sum_{n=1}^{\infty}a_{n}\cos{\frac{\pi nx_{2}}{L}} \right) \phi(x_{2})\\
&=I_{0}^{2}\left(\frac{L}{2}a_{0}^{2} + L\sum_{n=1}^{\infty}a_{n}^{2}\exp{\left(-\frac{Dn^{2}\pi^{2}t}{L^{2}}\right)}\right)
\end{aligned}
\end{equation}
as in the case with fitting membrane fluctuations, the autocorrelation function is a sum over several modes that decay with different relaxation times.  When we take the limit of $L$ to $\infty$ then the sum will be replaced by an integral and the standard autocorrelation function for FCS results.
\begin{figure}
    \centering
    \begin{subfigure}[b]{0.4\textwidth}
        \includegraphics[width=\textwidth]{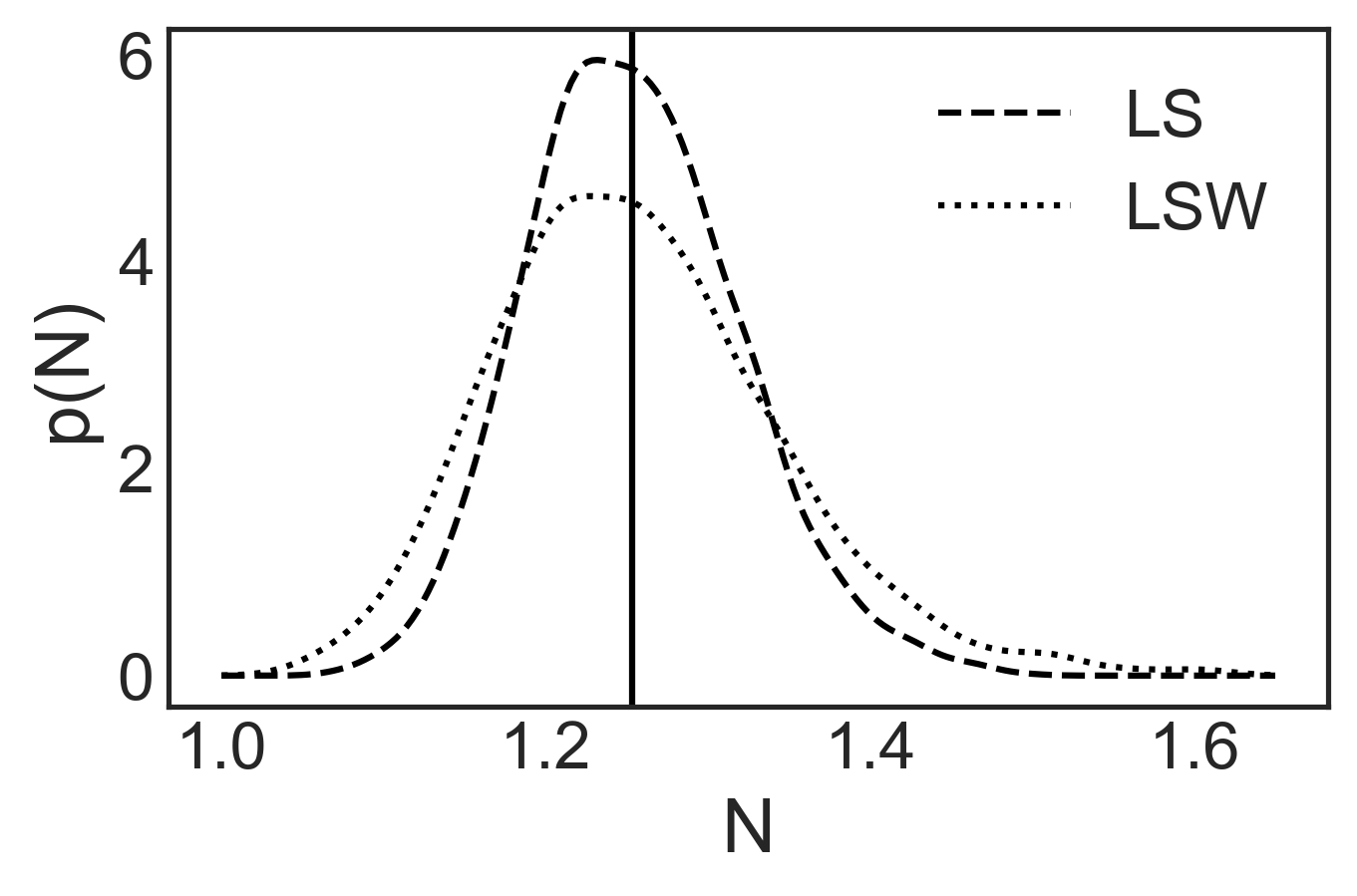}
        \caption{histogram of number of particles $N$}
        \label{fig:p_N}
    \end{subfigure}
    \qquad 
    \begin{subfigure}[b]{0.4\textwidth}
        \includegraphics[width=\textwidth]{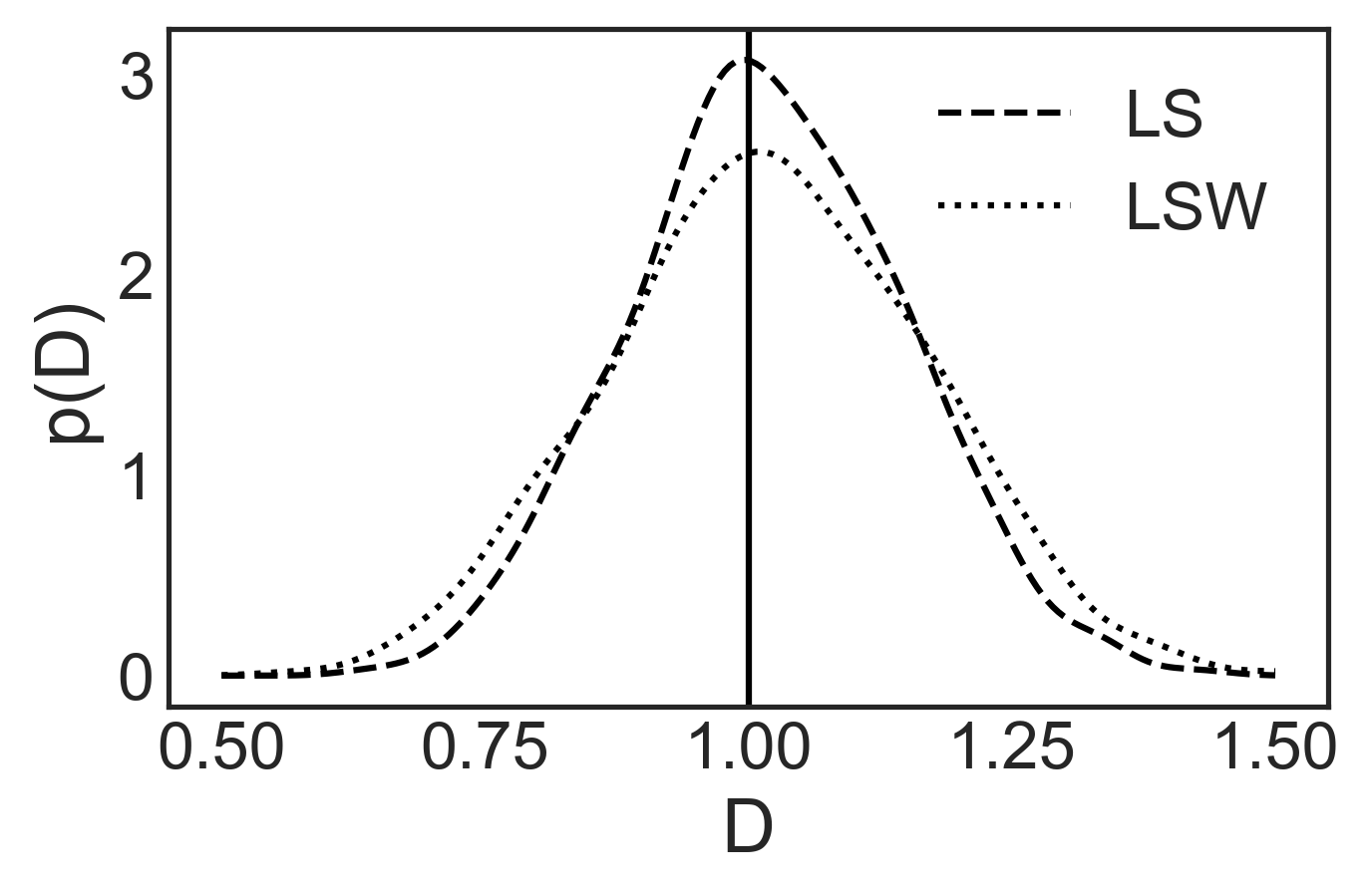}
        \caption{histogram of diffusion constant $D$}
        \label{fig:p_D}
    \end{subfigure}
    \caption{Comparison of Gaussian KDE probability distributions of $N$ and $D$ for traces of 100,000 points determined by least square fit with and without weights. }\label{fig:p_ND}
	\label{fig:FCSfig}
\end{figure}
Fig. \ref{fig:FCSfig} shows the simulation results.  Specifically, we used $\Delta t = 0.1$, $D=1$, $w=0.5$, $L=10$, and $N=20$.  We first randomly places the $N=20$ particles into our $(-L,L)$ box and then simulated 100,000 consecutive time points (corresponding to 160,000 diffusion times $\tau = w^{2}/4D$) during which all particles diffused freely while imposing reflective boundary conditions.  We then calculated the fluorescence intensity by adding the fluorescence contribution of each particle at each time, assuming that the illumination profile is described by a Gaussian eq. \ref{illpro} at $x=0$ with $w=0.5$.  At $\bar{c}=2L/N =1$ we would expect an average of $\sqrt{\pi/2}=1.25$ particles in the focus.  We simulated 500 runs, calculated the acf using FFT, and least-square fitted the theoretical expression for the autocorration function estimating $N$ and $D$.  As before, we fitted the autocorrelation functions assuming uniform standard deviation (LS) and fitting using appropritate weights (LSW).  From the distribution of fitted $D's$ we found that $D_{LS}=1.015\pm0.125$ and $D_{LSW}=1.017\pm0.148$, whereas $N_{LS}=1.26\pm0.06$ and $N_{LSW}=1.25\pm0.09$.  As before the least-square fits strongly underestimate the confidence interval as compared to the distribution: $\Delta D_{LS}=9.57\cdot 10^{-3}$, $\Delta D_{LSW}=1.15\cdot 10^{-2}$ and $\Delta N_{LS}=0.007$,$\Delta N_{LSW}=0.011$.  Here we underestimate the confidence interval by about a factor of 10 for both paramters.
From Fig. \ref{fig:FCSfig} one could argue that even though we overestimate our confidence using least-square fits, the mean values of the distributions still return the correct values.  if that were true, we could estimate the confidence simply by repeated measures.  To test this assuption, we simulated a shorter dataset with only 10,000 data points or 16,000 diffusion times (Fig. \ref{fig:FCSfig_short}).  In this case we observe that the probability distribution for the diffusion coefficient clearly deviates from the true mean of $D=1$.  In particular, we find that $D_{LS} = 1.15 \pm 0.012$ and $D_{LSW}=1.17 \pm 0.014$ using standard errors, placing the measured mean over 10 standard errors away from the true mean.  We anticipate that this deviation will increase as the length of the time series decreases.

This final example provides further evidence that least-square fits of autocorrelation functions are unreliable both in terms of an overestimation of confidence and a systematic deviation from the true mean.  Since FCS applications are moving more and more into dynamic biological systems, we anticipate the effects of least-square fitting are intensified due to the necessity of short time series.  This strongly suggests that we urgently need a probabilistic description of FCS similar to eq. \ref{OUp}, $p(I,t\left|I_{0},t_{0}\right.)$, which would allow us to express the likelihood function of an FCS intensity trace as function of all the model parameters.
\section{Conjugate Priors}
Our maximum likelihood solution currently assumes a constant prior for $A$ and $B$.  To introduce a reasonable prior we will first assume that the prior $p(B,A)=p(B)p(A)$ which means that the prior assumes independence of $B$ and $A$.  An appropriate prior for $p(B)$ is a beta distribution since it covers the interval $[0,1]$.  For example, for small $\Delta t$ it make sense to choose a beta distribution with $\alpha_{B} > 1$ and $\beta_{B} = 1$ to reflect the fact that we expect $B$ to be close to $1$.  The natural prior for $A$ is the inverse gamma distribution since it is a conjugate prior distribution.
The posterior probability distribution is then represented by:
\begin{eqnarray}
	&&p(B,A \left| \left\{x_i(t_i)\right\} \right. ) 
	\propto p(B)p(A)p\left( \left\{x_i(t_i)\right\} \left| B, A \right.\right)\nonumber\\
	&&= B^{\alpha_{B}-1}(1-B)^{\beta_{B}-1}
	\frac{1}{A^{\alpha_{IG}+1}}\exp \left( -\frac{\beta_{IG}}{A}\right)\nonumber\\
	&&\times\frac{1}{\sqrt {2 \pi A}^{N} }
	\frac{1}{{\sqrt {(1-B^{2})}^{(N-1)} }}
	\exp \left( -\frac{Q(B)}{2A}\right)
\end{eqnarray}
where $\alpha_{B}$ and $\beta_{B}$ are the shape parameters of the Beta distribution and $\alpha_{IG}$ and $\beta_{IG}$ for the inverse gamma distribution.
This changes the logarithm of the posterior $\Phi$ to:
\begin{eqnarray}
	\Phi =&& C +(\alpha_{B}-1)ln(B) + (\beta_{B}-1)ln(1-B)\nonumber\\
	&& - \frac{(N+2(\alpha_{IG}+1))}{2} ln(A) - \frac{N-1}{2}ln \left( 1-B^{2}\right)\nonumber\\
	&&-\frac{1}{2A}(Q(B)+2\beta_{IG})
\end{eqnarray}
this addition does change $A_{max}$ (see equation \ref{partialsigma}) to:
\begin{equation}
	A_{max} = \frac{Q(B_{max})+2\beta_{IG}}{N+2(\alpha_{IG}+1)}
	\label{priorA}
\end{equation}
While the derivative with respect to B is then:
\begin{equation}\label{partialB2}
	\frac{\partial}{\partial B}\Phi = 
	\frac{\alpha_{B}-1}{B}+\frac{\beta_{B}-1}{1-B}+
	\frac{(N-1)B}{1-B^{2}} -\frac{1}{2A}\frac{\partial}{\partial B}Q(B)\\
\end{equation}
again, we can combine eqs \ref{priorA} and \ref{partialB2} to calcualte $B_{max}$:
\begin{eqnarray}
	&&(Q(B_{max})+2\beta_{IG})\nonumber\\
	&&\times\left( \frac{\alpha_{B}-1}{B_{max}}+\frac{\beta_{B}-1}{1-B_{max}}+
	\frac{(N-1)B_{max}}{1-B_{max}^{2}}\right)\nonumber\\
	&&=\frac{N+2(\alpha_{IG}+1)}{2}\frac{\partial}{\partial B}Q(B)\mid_{B_{max}}
\end{eqnarray}
as was the case previously, $B_{max}$ can be calculated by finding the root of a polynomial in $B_{max}$ in the $[0,1]$ interval.

\section{Summary}
In the previous sections, we demonstrated that least-square fitting decaying time-autocorrelation functions is problematic.  We showed that even though the values of the least-square fit are in a reasonable range, the confidence intervals are not.  This may lead to an overconfidence in the measured data and may prevent the collection of more data, especially if the data is expensive to aquire.  In order to remedy this situation, we analytically solved the maximum likelihood expression for an Uhlenbeck-Ornstein process.  This solution, as compared to the least-square fit, does exhibit the correct confidence intervals.  Our analytic solution also identified an optimal sampling frequency for the relaxation rate $\tau$ which will allow to adaptively sample to obtain the highest accuracy given the least amount of data.  Unfortunately, our solution is only applicable to systems that display a single exponential decay time, as for example in dynamic light scattering of a single species.  When considering more complicated systems we need to develop more complicated models by combining many Uhlenbeck-Ornstein processes into a posteriors by convolution \cite{RN43}.  Another possibility is to use Markov-Chain-Monte-Carlo methods to evaluate these systems by randomly sampling from the posterior \cite{RN46}.  Our analysis makes clear that conditional probability approaches are urgently needed for experimental methods such as FCS in order to perform precision experiments.  Even though these techniques are often used for biological systems that typically exhibit a large variability, we would argue that only by knowing the correct confidence intervals, we can properly characterize biological variance.  As quantiative scientists we should strive for the best precision attainable.

\begin{acknowledgments}
We wish to acknowledge funding by the NSF (DMR Award
1106044), the NIH (5R21DA03846702), and discussions with Ken Dill, Corey Weistuch, Don Lamb and Joachim R\"adler.
\end{acknowledgments}

\bibliography{langevin_pre}

\end{document}